\documentclass[12pt]{article}
\usepackage{amstex}
\usepackage{psfig}
\makeatletter

\newcommand{\mrm}[1]{\mathrm{#1}}

\newcommand{\JP}{{\mrm{J}}/\psi}
\renewcommand{\b}{\mrm{b}}
\renewcommand{\c}{\mrm{c}}
\renewcommand{\d}{\mrm{d}}
\newcommand{\e}{\mrm{e}}
\newcommand{\g}{\mrm{g}}

\newcommand{\q}{\mrm{q}}

\newcommand{\mc}{m_{\c}}
\newcommand{\Mc}{M_{\c}}
\newcommand{\bbar}{\overline{\mrm{b}}}
\newcommand{\cbar}{\overline{\mrm{c}}}

\newcommand{\qbar}{\overline{\mrm{q}}}

\newcommand{\trv}{_{\perp}}
\newcommand{\BR}{\mbox{BR}}

\newcommand{\LQCD}{\Lambda_{\rm{QCD}}}

\newcommand{\sla}{\hspace*{-0.20cm}/}
\newcommand{\lessim}{\raisebox{-0.8mm}%
{\hspace{1mm}$\stackrel{<}{\sim}$\hspace{1mm}}}
\newcommand{\alphas}{\alpha_{\mrm{s}}}

\newcommand{\as}{{\mrm{as}}}

\newcommand{\fpi}{f_{\pi}}
\newcommand{\da}{distribution amplitude}
\newcommand{\das}{distribution amplitudes}
{\end{list}}
\newcounter{enumct}

\newlength{\abstwidth}
\begin{document}
 
%set sloppy attitude to line breaks
\sloppy

\renewcommand{\arraystretch}{1.5}

\pagestyle{empty}

\begin{flushright}
CERN--TH/96--266  \\
WU B 96-28
\end{flushright}
 
\vspace{\fill}
 
\begin{center}
{\LARGE\bf 
%The decay of P-wave charmonia into two pions
Colour-octet contributions to\\[2ex] exclusive charmonium decays
}\\[10mm]
{\Large Jan Bolz$^a$, Peter Kroll} \\[3mm]
{\it Fachbereich Physik,}\\[1mm]
{\it University of Wuppertal, Germany}\\[1mm]
{ E-mail: kroll@wptu15.physik.uni-wuppertal.de}\\[2ex]
{\large and} \\[2ex]
{\Large Gerhard A. Schuler$^b$} \\[3mm]
{\it Theory Division, CERN,} \\[1mm]
{\it CH-1211 Geneva 23, Switzerland}\\[1mm]
{ E-mail: Gerhard.Schuler@cern.ch}
\end{center}
 
\vspace{\fill}
 
\begin{center}
{\bf Abstract}\\[2ex]
\begin{minipage}{\abstwidth}
We investigate the theoretical uncertainties of $P$-wave charmonium 
decays into two pions, $\chi_{\c J}\rightarrow \pi^+\pi^-$, $\pi^0\pi^0$. 
Constraining the pion distribution-amplitude from the recent precise 
data on $F_{\pi\gamma}(Q^2)$, we find an order-of-magnitude discrepancy 
between data and prediction. The disagreement persists even after 
inclusion of transverse degrees of freedom and Sudakov suppressions. 
We propose the colour-octet mechanism as the solution to the puzzle.
The colour-octet decay contribution arising from the higher Fock component
$|\c \cbar \g \rangle$ of the $\chi_{\c J}$ wave function is actually
not power suppressed with respect to the usual colour-singlet decay 
arising from the dominant $|\c \cbar \rangle$  Fock state.
An explicit calculation yields an agreement with the data for a very 
reasonable value for the single extra non-perturbative parameter.
\end{minipage}
\end{center}

\vspace{\fill}
\noindent
\rule{60mm}{0.4mm}

\vspace{1mm} \noindent
${}^a$ Supported by Deutsche Forschungsgemeinschaft.\\
${}^b$ Heisenberg Fellow.

\vspace{10mm}\noindent
CERN--TH/96--266 \\
September 1996

\clearpage
\pagestyle{plain}
\setcounter{page}{1} 

%%%%%%%%%%%%%%%%%%%%%%%%%%%%%%%%%%%%%%%%%%%%%%%%%%%%%%%%%%%%%%%%%%
%\section{Introduction}
%%%%%%%%%%%%%%%%%%%%%%%%%%%%%%%%%%%%%%%%%%%%%%%%%%%%%%%%%%%%%%%%%%
Up to very recently, hard exclusive reactions involving pions could 
be believed to be successfully described within the hard-scattering 
approach. The reactions to be mentioned are the pion--photon
transition form factor, the pion electromagnetic form factor, 
pion-pair production by two photons, as well as the two-pion decays 
of the charmonium states $\JP$, $\chi_{\c 0}$, and $\chi_{\c 2}$. 
The hard-scattering approach (HSA) provides the framework of 
calculations of exclusive reactions at large momentum transfers 
\cite{Brodsky80}: a full amplitude factors into two parts, 
a hard-scattering amplitude $T_{H}$, calculable in 
perturbative QCD, and parton-distribution amplitudes $\Phi_H$ for 
each hadron $H$. The amplitude $T_{H}$ describes the scattering 
of clusters of collinear partons from the hadron. 
The leading-twist contribution, i.e.\ the one with the weakest 
fall-off with $Q$, the typical momentum transferred in the process, 
is given by valence-parton scatterings only. Hence the only 
non-perturbative input required are the probability amplitudes 
(or quark \das)
$\Phi_H(x_i,Q)$ for finding valence quarks in the hadron, each 
carrying some fraction $x_i$ of the hadron's momentum. 

In the case of mesons, the leading Fock state is the 
$| \q\qbar\ \rangle$ state, which, for the pion in the isotopic 
limit, can be described by a symmetric wave function 
$\Phi_{\pi}(x) = \Phi_{\pi}(1-x)$. Upon expansion over Gegenbauer 
polynomials $C_n^{(3/2)}$, the \da\ at a scale $\mu_F$ can be 
characterized by non-perturbative coefficients $B_n$ (see \cite{Brodsky80} 
and references therein):
\begin{equation}
  \Phi_\pi(x,\mu_F) = \frac{f_\pi}{2 \sqrt{6}}\, \phi_{\as}(x)\, 
  \left[ 1 + \sum_{n=2,4,\ldots}^{\infty}\, B_n 
   \left( \frac{\alpha_s(\mu_F) }{\alpha_s(\mu_0) } \right)^{\gamma_n}
  \, C_n^{(3/2)}(2x-1) \right]
\ .
\label{phiex}
\end{equation}
Here $f_\pi$ is the pion decay constant, $f_\pi = 130.7\,$MeV,
$\alpha_s$ the strong coupling constant, and $\mu_0$ a hadronic scale,
$0.5 \lessim \mu_0 \lessim 1\,$GeV.
Since the $\gamma_n$ in (\ref{phiex}) are positive fractional numbers 
increasing with $n$, higher-order terms are gradually suppressed and
any \da\ evolves into $\phi_{\as}(x) = 6x(1-x)$ 
asymptotically, i.e.\ for $\ln(Q^2/\LQCD^2) \rightarrow \infty$. 

Lowest-order perturbative QCD (pQCD) calculations 
in leading-twist of the above-mentioned reactions 
involving pions were all in reasonable agreement with data, provided the 
quark \da\  in the pion is strongly end-point-concentrated
(in sharp contrast to the asymptotic \da). 
Such a \da\ is given by the Chernyak--Zhitnitsky (CZ) \cite{CZ82} one 
defined by $B_2 = 2/3$ and $B_n=0$ for $n>2$. So far, 
there existed basically only one exception, namely the pion form factor
in the time-like region: the HSA prediction with the CZ \da\ fails by 
about a factor of $2$ as compared with the data. 
The experimental information on the time-like form
factor comes from two sources, 
$\e^+\e^-\to \pi^+\pi^-$ and $\JP \to\pi^+\pi^-$, 
which provides, to a very good approximation, the form factor at
$s=M^2_{\JP}$. Although the $\e^+e^-$ annihilation data of Ref.~\cite{Bol75}
suffer from low statistics, they agree very well with the result obtained
from the $\JP$ decay.

Meanwhile, the situation has changed drastically. On the experimental side,
very precise new data for fairly large momentum transfer \cite{CLEO} 
have shown that $F_{\pi\gamma}(Q^2)$ is smaller than predicted from
the CZ \da. On the theoretical side, the HSA has been modified 
through the incorporation of transverse degrees of freedom and Sudakov
suppressions \cite{Bot89,Li92}. This more refined treatment (termed
the ``modified HSA'') allows the calculation of the genuinely perturbative 
contribution self-consistently, in the sense that the bulk of the 
perturbative contribution is accumulated in regions of reasonably 
small values of the strong coupling constant $\alphas$. This approach
hence overcomes arguments \cite{Isg89} against the applicability of 
the standard HSA (sHSA: the lowest-order pQCD approach in the
collinear approximation using valence Fock states only) 
to experimentally accessible regions of momentum transfers. It is to be 
stressed that the effects of the transverse degrees of freedom taken 
into account in the modified HSA (mHSA) represent soft contributions of 
higher-twist type. Still, mHSA calculations are restricted to the 
dominant (valence) Fock state.

A detailed analysis of the photon--pion transition form factor 
$F_{\pi\gamma}(Q^2)$ within the mHSA \cite{Jak96,Rau96} shows that 
the pion \da\ is close to the asymptotic form $\phi_{\as}(x)$. In order
to give a quantitative estimate of the allowed deviations from the
asymptotic \da, one may assume that $B_2$ is the only non-zero
expansion coefficient in (\ref{phiex}). The truncated series suffices to
parametrize small deviations. Moreover, it has the advantage of
interpolating smoothly between the asymptotic \da\ and the frequently
used CZ \da. For large momentum transfer our assumption is
justified by the properties of the anomalous dimensions $\gamma_n$. 
Quantitatively, choosing $\mu_0=0.5\,$GeV, a best fit to the
$F_{\pi\gamma}$ data above $1\,$GeV$^2$ yields
\begin{eqnarray}
B_2(\mu_0) & =  -0.006 \pm 0.014 \qquad & {\mrm{mHSA}}
\nonumber\\
           & =  -0.39~ \pm 0.05~ \qquad & {\mrm{sHSA}}
\ .
\label{eq:B2values}
\end{eqnarray} 

The non-perturbative pion \da\ (i.e.\ the coefficient $B_2$) is, in fact, 
best extracted from $F_{\pi\gamma}(Q^2)$, since the 
pseudoscalar meson--photon transition form factors are pure QED 
processes with QCD corrections of only 10--20\% and higher Fock-state 
contributions suppressed by powers of $\alpha_s/Q^2$ and expected to
be small too \cite{Gor89}. Uncertainties due to these effects may be
absorbed into a theoretical error of $B_2$, which we generously estimate 
to amount to $\pm0.1$ for the mHSA and $\pm 0.39$ for the sHSA.
However, with a \da\ close to the asymptotic
form, calculations for the other processes involving pions disagree 
strongly with data, at least when calculated in the collinear 
approximation. In this paper we shall investigate in detail the 
$\chi_{\c J}$ decays into two pions,  
by calculating the decay rates and branching ratios in the modified
HSA, using the information on the process-independent expansion coefficient
$B_2$ obtained from the $\pi\gamma$ form factor.

Using the method proposed by Duncan and Mueller \cite{Dun80}, 
the decay rate calculated in the sHSA can be written as
\begin{equation}
  \Gamma[ \chi_{\c J} \rightarrow \pi^+\pi^-] = 
      f_{\pi}^4\, \frac{1}{\mc^8}\, |R'_P(0)|^2\, \alpha_s(\mu_R^{excl})^4\, 
  \left| a_J + b_J\, B_2(\mu_0) + c_J\, B_2(\mu_0)^2 \right|^2
\ ,
\label{decayzero}
\end{equation}
where $a_J$, $b_J$, and $c_J$ are (analytically) calculable real numbers,
see table~\ref{tab:Cicoeffs}.
%%%%%%%%%%%%%%%%%%%%%%%%%%%%%%%%%%%%%%%%%%%%%%%%%%%%%%%%%%%%%%%%%%%%%%%%%
\begin{table}
\begin{center}
\begin{tabular}{|l||l|l|l|} \hline
 Approach   & $a_0$  & $b_0$  & $c_0$ \\ \hline 
  sHSA& $\phantom{-}19.60$  & $37.43$  & $26.23$ \\ \hline
  mHSA& $ -14.67 + 23.36\,\imath$ & $15.36 + 73.22\,\imath$ &
        $27.18 + 26.39\,\imath$ \\ \hline\hline
      & $a_2$  & $b_2$ & $c_2$  \\ \hline
  sHSA& $\phantom{-}4.589$ & $8.479$ & $7.719$ \\ \hline
  mHSA& $-3.420 +  5.141\,\imath$ & $ 6.681 + 13.42\,\imath$ & 
        $ 5.744 +  1.384\,\imath$  \\ \hline\hline
\end{tabular}
\end{center}
\caption[dummy1]{Coefficients defined in (\ref{decayzero})
  obtained within the standard HSA (sHSA) for $\mu_R^{excl}=\mc=1.5\,$GeV
  and coefficients (\ref{decayone}) of the modified HSA (mHSA).}
\label{tab:Cicoeffs}
\end{table} 
The evolution factor $[\ln(\mu_F^2/\LQCD^2)/\ln(\mu_0^2/\LQCD^2)]^{\gamma_2}$ 
(see (\ref{phiex})) is absorbed into the coefficients $b_J$ and $c_J$ 
(quadratically in the latter case), and $\gamma_2=50/81$. $\mu_R^{excl}$ 
is an appropriate renormalization scale, which is of the order of the
charm-quark mass $\mc$ \cite{Brodsky80,Bar81,Bod94}. In the sHSA 
analysis, it is customary to identify factorization scale $\mu_F$ and
renormalization scale. Finally, $R'_P(0)$
denotes the derivative of the non-relativistic $\c\cbar$ wave function
at the origin (in coordinate space) appropriate for the dominant Fock state 
of the $\chi_{\c J}$, a $\c\cbar$ pair in a colour-singlet state 
with quantum numbers ${}^{2S+1}L_J = {}^3P_J$.

%%%%%%%%%%%%%%%%%%%%%%%%%%%%%%%%%%%%%%%%%%%%%%%%%%%%%%%%%%%%%%%%%%%%%%%%%
\begin{table}
\begin{center}
\begin{tabular}{|r|r|r|r|r|r|r|} \hline
  $B_2$ & $\mc$ & $\LQCD$ 
  & \multicolumn{2}{|c|}{$\Gamma(\chi_{\c J} \to \pi^+\pi^-)\,$[keV]  }
  & \multicolumn{2}{|c|}{$\BR(\chi_{\c J} \to \pi^+\pi^-)\,$[\%]}
\\ \hline
 & [GeV] & [GeV] & $J=0$  & $J=2$  & $J=0$  & $J=2$ 
\\ \hline\hline
  \multicolumn{7}{|c|}{Standard HSA}
\\ \hline
 $-0.39$ & $1.5$ & $0.2$ & $0.872$ & $0.065$  & $0.011$ & $0.003$ 
\\ \hline
 $0$ & $1.35$ & $0.25$ & $15.3$ &  $0.841$ & $0.113$ & $0.020$
\\ \hline\hline
  \multicolumn{7}{|c|}{Modified HSA}
\\ \hline
  $0$ & $1.5$ & $0.2$  
  &   8.22  &  0.41  &  0.102  &  0.017  
\\ \hline
  $0.1$ & $1.5$ & $0.2$   
  &   12.13  &  0.53  &  0.151  &  0.021  
\\ \hline
  $-0.1$ & $1.5$ & $0.2$   
  &   5.61  &  0.33  &  0.070  &  0.013  
\\ \hline
%--------------------------------------------------
%     mc = 1.8 GeV and 1.35 GeV 
%--------------------------------------------------
  $0$ & $1.8$ & $0.2$   
  &  2.54  &  0.12  &  0.050  &  0.008  
\\ \hline
  $0$ & $1.35$ & $0.2$   
  & 15.3   & 0.78   & 0.144   &  0.024  
\\ \hline
%--------------------------------------------------
%     Lambda = 150 MeV 
%--------------------------------------------------
  $0$ & $1.5$ & $0.15$   
  &   4.34  &  0.21  &  0.071  &  0.011  
\\ \hline
%--------------------------------------------------
%     Lambda = 250 MeV 
%--------------------------------------------------
  $0$ & $1.5$ & $0.25$   
  &  13.1  &  0.68  &  0.128  &  0.022  
\\ \hline\hline
  \multicolumn{7}{|c|}{Standard HSA plus colour-octet contributions}
\\   \hline
$0$ & $1.5$ & $0.2$ & $49.85$ & $3.54$ & $0.36$ & $0.177$
\\ \hline\hline
  \multicolumn{7}{|c|}{Experiment}
\\   \hline
\multicolumn{3}{|c|}{PDG\cite{pdg94}}
       &  $105 \pm 30$   &  $3.8 \pm 2.0$   
       &  $0.75 \pm 0.21$    &  $0.19 \pm 0.10$  
\\   \hline
\multicolumn{3}{|c|}{BES\cite{BEBC}}
       &  $62.3 \pm 17.3$   &  $3.04 \pm 0.73$   
       &  $0.427 \pm 0.064$    &  $0.152 \pm 0.034$  
\\   \hline
\end{tabular}
\end{center}
\caption[dummy2]{Decay widths and branching ratios of $\chi_{\c J} \to
  \pi^+\pi^-$
  for various choices of the parameters compared with experimental
  data. The standard HSA results are obtained with 
  $\mu_R^{excl} =\mc$, the branching ratios are evaluated with
  $\mu_R^{incl} =\mc$ and $F_1 = 1$.
  \label{tab:results}
}
\end{table} 
In the following we will choose as central values 
$R'_P(0)= 0.22$ GeV$^{5/2}$ and $\mc = 1.5\,$GeV, which is
consistent with a global fit of charmonium parameters \cite{MaP95} as
well as results for charmonium radii from potential models\cite{BuT81}. 
With this choice, $\mu_R=\mc$, and $B_2(\mu_0)=-0.39$ 
(see (\ref{eq:B2values}))
the sHSA results for the
$\chi_{c0(2)} \to \pi^+\pi^-$ decay widths are 
almost two orders of magnitude below the experimental data, 
see table~\ref{tab:results}. To assess the uncertainty
of the result we vary the charm-quark mass between $1.35\,$GeV
and $1.8\,$GeV. Of course, the value of $R'_P(0)$ has to be
adjusted accordingly: using the well-known scaling properties of 
quarkonium potential models \cite{Quigg79} we take 
$R'_P(0) = 0.274\,$GeV$^{5/2}$ for $\mc = 1.8\,$GeV 
and $R'_P(0) = 0.194\,$GeV$^{5/2}$ for $\mc = 1.35\,$GeV. 
However, even when stretching all parameters to their extreme values,
the predictions stay a factor $3$--$6$ below the data. 
Note that\footnote{We calculate $\alpha_s$ in the one-loop approximation 
with $\LQCD = 200\,$MeV and $n_f = 4$.} $\alpha_s(\mc)=0.447$ for
$\mc=1.35\,$GeV and $\LQCD=0.25\,$GeV. 

\begin{figure}
\[
 \psfig{figure=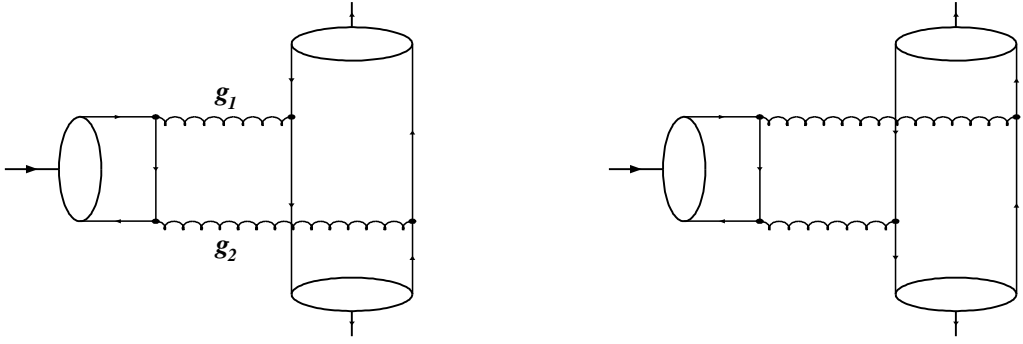,%
       bbllx=50pt,bblly=40pt,bburx=750pt,bbury=285pt,%
       height=5cm,clip=} \]
 \caption[dummy0]{Feynman diagrams for the colour-singlet decay
  $\chi_{\c J} \rightarrow \pi\, \pi$ ($J=0,2$).  
  \label{fig:col1graph} }
\end{figure}

The starting point of the calculation of the $\chi_{\c J}$ decays within
the modified HSA is the convolution formula\begin{multline}
  M(\chi_{\c J} \to \pi^+\pi^-) = \frac{32\,\sqrt{2}\,\pi^{3/2}\,R'_P(0)}
      {3\sqrt{3}\,\mc^{7/2}} \sigma_J\, \\
      \int_0^1 \d x \d y \, \int \frac{\d^2 {\bf b}}{(4\pi)^2} \,
      \hat \Psi_{\pi}^{\ast}(y,{\bf b})\,\hat T_{HJ}(x,y,{\bf b})\,
      \hat \Psi_{\pi}(x,{\bf b})\,\exp[-S(x,y,{\bf b},t_1,t_2)]
\ ,
  \label{Mb}
\end{multline}
which adapts the methods proposed by Botts and Sterman \cite{Bot89} to
our case (\,$\sigma_0=1$; $\sigma_2=\sqrt{3}/2$\,). Note that we work in the transverse 
coordinate space (the quark--antiquark separation $\vec{b}$ is
canonically conjugated to the usual transverse momentum $\vec{k}_{\perp}$). 
$\hat T_{HJ}$ is the Fourier-transformed hard-scattering amplitude
to be calculated from the graphs shown in Fig.~\ref{fig:col1graph}:
\begin{equation}
  \hat{T}_{HJ}(x,y,{\bf k}\trv\!-\!{\bf k'}\trv) 
    = \frac{\alphas(t_1)\,\alphas(t_2)}
           {(g_1^2 + \imath\epsilon)\,(g_2^2 + \imath\epsilon)\,N} \,
      \left( 1 + \frac{(-2)^{J/2}}{2}\frac{(x-y)^2}{N} \right) \: ,
  \label{THkt}
\end{equation}
and with ${\bf k} = ({\bf k}\trv\!-\!{\bf k'}\trv)/2 \mc$
\begin{eqnarray}
  N & = &  x (1-y) + (1-x) y + 2\,{\bf k}^2 \, ,\nonumber \\
  g_1^2 & = &  x y - {\bf k}^2 \, ,\nonumber \\
  g_2^2 & = & (1-x)(1-y) - {\bf k}^2 . 
  \label{Ngi}
\end{eqnarray}

In analogy to previous applications of the modified 
perturbative approach, the Sudakov exponent is given by 
\begin{eqnarray}
  S(x,y,{\bf b},t_1,t_2) & = & s(x,b,2 \mc) + s(1-x,b,2 \mc)
                             + s(y,b,2 \mc) + s(1-y,b,2 \mc)
                             \nonumber \\
                         & - & \frac{4}{\beta} \log
         \frac{\log(t_1/\LQCD)\log(t_2/\LQCD)}{\log(1/(b \LQCD))^2} \,,
  \label{sudexp}
\end{eqnarray}
where the function $s(x,b,Q)$ can be found, for instance, in
Ref.~\cite{DJK95}.  

The renormalization scales $t_i = \max(2 \sqrt{x_i y_i} \mc,1/b)$
($i=1,2$) appearing in $\alphas$ and in the Sudakov
exponent are determined by the virtualities of the 
intermediate gluons, which depend non-trivially on the integration
variables. This choice of the renormalization
scale avoids large logs from higher-order pQCD. The 
factorization scale is given by the quark--antiquark separation
$\vec{b}$, $\mu_F = 1/b \equiv 1/|\vec{b}|$. 
The ratio $1/b$ marks the interface
between non-perturbatively soft momenta, which are implicitly accounted
for in the pion wave function $\Psi_{\pi}$, and the contributions
{from} semi-hard gluons, incorporated in a perturbative way in the 
Sudakov factor. 
 
The last object appearing in (\ref{Mb}) is the full (soft) wave function 
of the pion describing also the dependence on the transverse momentum. 
Following \cite{Jak93} we make the ansatz
\begin{eqnarray}
  \hat \Psi_{\pi}(x,b;\mu_F) & = & \frac{f_{\pi}}{2\,\sqrt{6}} \, 
  \Phi_{\pi}(x,\mu_F) \, \hat\Sigma_{\pi}(x,b) \, , 
\nonumber \\
  \hat\Sigma_{\pi}(x,b)    & = &  4\pi \, 
        \exp\left[-\frac{x (1-x) b^2}{4\,a_{\pi}^2}\right] \:.
\label{eq:ansatz}
\end{eqnarray}  
Note, however, that $a_\pi$ is not a free parameter since it is fixed 
{from} $\pi^0 \rightarrow \gamma\gamma$ \cite{Bro83}. That constraint leads
to the closed formula 
$1/a_\pi^2 =  8\, (1+B_2)\ \pi^2\  f_{\pi}^2$. 
This additional
$B_2$ dependence is taken into account in our calculation by expanding
the Gaussian in $\hat\Sigma$ over $B_2$. For the asymptotic form of
the wave function ($B_2=0$) the transverse size parameter $a_{\pi}$
takes the value 0.861 GeV$^{-1}$. 

In terms of the amplitude (\ref{Mb}) the decay widths are given by
\begin{eqnarray}
  \label{width}
    \Gamma(\chi_{c0} \to \pi^+\pi^-) & = & \frac{1}{32\,\pi\,\mc}\:
    \left| M(\chi_{c0} \to \pi^+\pi^-)\right|^2 \label {Gamma0} \nonumber\\
  \Gamma(\chi_{c2} \to \pi^+\pi^-) & = & \frac{1}{240\,\pi\,\mc}\:
    \left| M(\chi_{c2} \to \pi^+\pi^-)\right|^2 
\ .
\label {Gamma2} 
\end{eqnarray}
The final results in the modified HSA, tables~\ref{tab:Cicoeffs} 
and~\ref{tab:results} can be obtained numerically only,
but can be cast into a form similar to (\ref{decayzero})
\begin{equation}
  \Gamma[ \chi_{\c J} \rightarrow \pi^+\pi^-] = 
      f_{\pi}^4\, \frac{1}{\mc^8}\, |R'_P(0)|^2\, \alpha_s(\mc)^4\, 
  \left| a_J + b_J\, B_2(\mu_0) + c_J\, B_2(\mu_0)^2 \right|^2
\ .
\label{decayone}
\end{equation}
The coefficients $a_J$, $b_J$, and $c_J$ are now complex-valued.
It is still convenient to divide out the fourth power of 
$\alphas$ at the fixed scale $\mc$ in (\ref{decayone}), since the main 
effect of the strong coupling is thus made explicit. 
The actual effective renormalization scale  $\mu_R$ in the 
modified HSA differs from $\mc$ and depends on $B_2$. We find 
\begin{equation}
\begin{tabular}{rlcrll}
$\mu_R^{\rm eff}$ & $= 1.15\,$GeV & \quad , \quad 
   & $\alpha_s^{\rm eff}$ & $= 0.43$ & ($B_2=0$)
\\
                  & $= 0.85\,$GeV & \quad , \quad 
   &                      & $= 0.52$ & ($B_2 = 2/3$) \ .
\end{tabular}
\label{alsval}
\end{equation}
Note that part of the increase of the decay widths compared to the 
sHSA results follows from the larger values of the
strong coupling constant
$\alphas$, cf.\ (\ref{alsval}) with $\alphas(\mc^2) = 0.374$. 

The calculation of the charmonium decay is theoretically self-consistent 
in so far as only 2\% (20\%) for $B_2=0$ ($B_2=2/3$) come from soft 
regions where the use of pQCD is inconsistent. The soft region is defined 
by $\alphas(t_1^2)\, \alpha_s(t_2^2) > 0.5$.

Uncertainties in our predictions of the $\chi_{\c}$ 
decay rates into pions arise from uncertainties in the values 
of $B_2$, $\LQCD$, $R'_P(0)$ and the charm-quark mass $\mc$. 
In order to show the 
dependence on $\mc$ we give our results in table~\ref{tab:results} 
for the above three choices of $\mc$
(with $R'_P(0)$ adjusted accordingly).
We also show the dependences on $\LQCD$ and $B_2$, where we stretch 
$B_2$ up to the maximal value allowed by the analysis of the photon--pion 
transition form factor, see above.
As shown in table~\ref{tab:results} the predicted decay widths 
for both $\chi_{\c 0}$ and $\chi_{\c 2}$ are well below the data, even if the
parameters are pushed to their extreme values. 

In order to eliminate the dependence on $R'_P(0)$ and 
reduce the one on $\mc$ one can consider the branching ratios 
into $\pi^+\pi^-$ rather than the absolute widths. 
When normalising to the total width we correct for
the fraction $\rho_J$ of $\chi_{\c J}$ decays that goes into 
light hadrons (rather than decaying electromagnetically or into lighter 
charmonium states, and which can be determined from data).
The branching ratio has a form similar to (\ref{decayone}):
\begin{equation}
  \BR[ \chi_{\c J} \rightarrow \pi^+\pi^-] = F_1\,
   \frac{ \alphas(\mc)^4 }{ \alphas(\mu_{R}^{incl})^2 }\,
      \frac{ f_{\pi}^4 } {\mc^4}\, 
    d_J\,
  \left| a_J + b_J\ B_2(\mu_0) + c_J\ B_2(\mu_0)^2 \right|^2
\ ,
\label{branchratio}
\end{equation}
where $d_0=\rho_0/6$, $d_2=5\rho_2/8$, and $F_1=1$ (see below).
The results for the branching ratios, see 
table~\ref{tab:results}, are indeed less sensitive to the parameter
choice (and independent of $R'_P(0)$), and confirm our conclusion
based on the decay widths: 
The calculation based on the assumption that the $\chi_{\c J}$ is a pure
$\c\cbar$ state,
is not sufficient to explain the observed rates. The necessary 
corrections would have to be larger than the leading terms. A new 
mechanism is therefore called for.

In the quark-potential model, any charmonium state 
is a pure $\c\cbar$ state, where the quarks have angular momentum $L$
and spin $S$ such that these match the quantum numbers by which any meson
is characterized in QCD, total spin $J$,
parity $P$ and charge conjugation $C$: $\bf {J}= {\bf L} + {\bf S}$, 
$P = (-1)^{L+1}$, $C = (-1)^{L+S}$. The $\chi_{\c J}$ states are therefore 
made out of ${}^3P_J$ $\c\cbar$ states (in the spectroscopic notation
${}^{(2S+1)}L_J$), which, obviously, are in colour-singlet ($c=1$) states.
Although the HSA acknowledges higher Fock components 
in the meson's wave function, these do not contribute in 
either sHSA or mHSA calculations since their contributions are assumed
to be suppressed by powers of  the hard scale \cite{Brodsky80}.

Recently, the importance of higher Fock states in understanding the 
production and the {\em inclusive} decays of charmonia has been pointed 
out \cite{Bod94}. The heavy-quark mass allows for a systematic expansion 
of both the quarkonium state and the hard, short-distance process and, 
hence, of the inclusive decay rate or the production cross section. The
expansion parameter is provided by the velocity $v$ of the heavy quark
inside the meson. In the case of the $\chi_{\c J}$, the Fock-state
expansion starts as
\begin{equation}
  |\chi_{\c J}\rangle = O(1)\, |\c\cbar_1({}^3P_J)\rangle
                      + O(v)\, |\c\cbar_8({}^3S_1)\, \g \rangle
                      + O(v^2) \,
\label{Fockexpansion}
\end{equation}
where the subscript $c$ specifies whether the $\c\cbar_c$ is in a 
colour-singlet ($c=1$) or colour-octet ($c=8$) state. 

The crucial observation is now that, for inclusive decays 
(and also production rates), both states in (\ref{Fockexpansion}) contribute
at the same order in $v$ and, hence, the inclusion of the 
``octet mechanism'', i.e.\ the contribution from the 
$|\c\cbar_8\g\rangle$ state, is necessary for a consistent description.
(Without its inclusion the factorization of the decay width into long-
and short-distance factors is spoiled by the presence of infrared-sensitive
logarithms.) In simple terms this can be understood
by realizing that the annihilation of the $\c\cbar$ pair of the higher Fock 
state into light hadrons is an $S$-wave process compared
to the $P$-wave annihilation of the leading Fock state. 
The inclusion of the colour-octet contribution to the total 
$\chi_{\c J}$ width does, however, worsen the comparison of
(\ref{branchratio}) with data since the fraction $F_1$ of the total 
hadronic width, which proceeds through the $c=1$ contribution, is less 
than unity. 

Here we propose the inclusion of contributions from the
$|\c\cbar_8({}^3S_1)\g\rangle$ Fock state to the {\em exclusive} 
$\chi_{\c J}$ decays as the solution to the failure of the HSA.
We shall show that this colour-octet contribution is indeed 
not suppressed by powers of $1/\mc$ compared with 
the conventional, colour-singlet contribution. 
Let us first, however, discuss the problem of colour conservation. This is
a new feature, specific to exclusive decays. 

The obvious solution to colour conservation seems
to be to demand one of the final-state pions to be also in a higher
Fock state, i.e.\ to consider $|\c\cbar\g\rangle \rightarrow 
|\q\qbar\g\rangle + |\q\qbar\rangle$. 
The hard process consists of the Feynman diagrams
of Fig.~\ref{fig:col1graph} plus the diagram in which $\c$ and $\cbar$
annihilate into a single gluon, which, in turn, makes a $\q\qbar$ pair.
Yet, we disregard this possibility.
The reason has to do with the importance of higher Fock states
in the pion. Since the $\q\qbar$ pair of the dominant Fock state 
is in a ${}^1S_0$ state, it is in a ${}^1P_1$ state for the next-higher
state. The first $S$-wave state $|\q\qbar_8({}^3S_1)\g\rangle$ 
is reached at $O(v^2)$. Contributions from higher Fock states 
of the pion therefore seem to be suppressed.

\begin{figure}
 \unitlength 1mm
 \begin{picture}(160,100)
   \put( 5,30){\psfig{figure=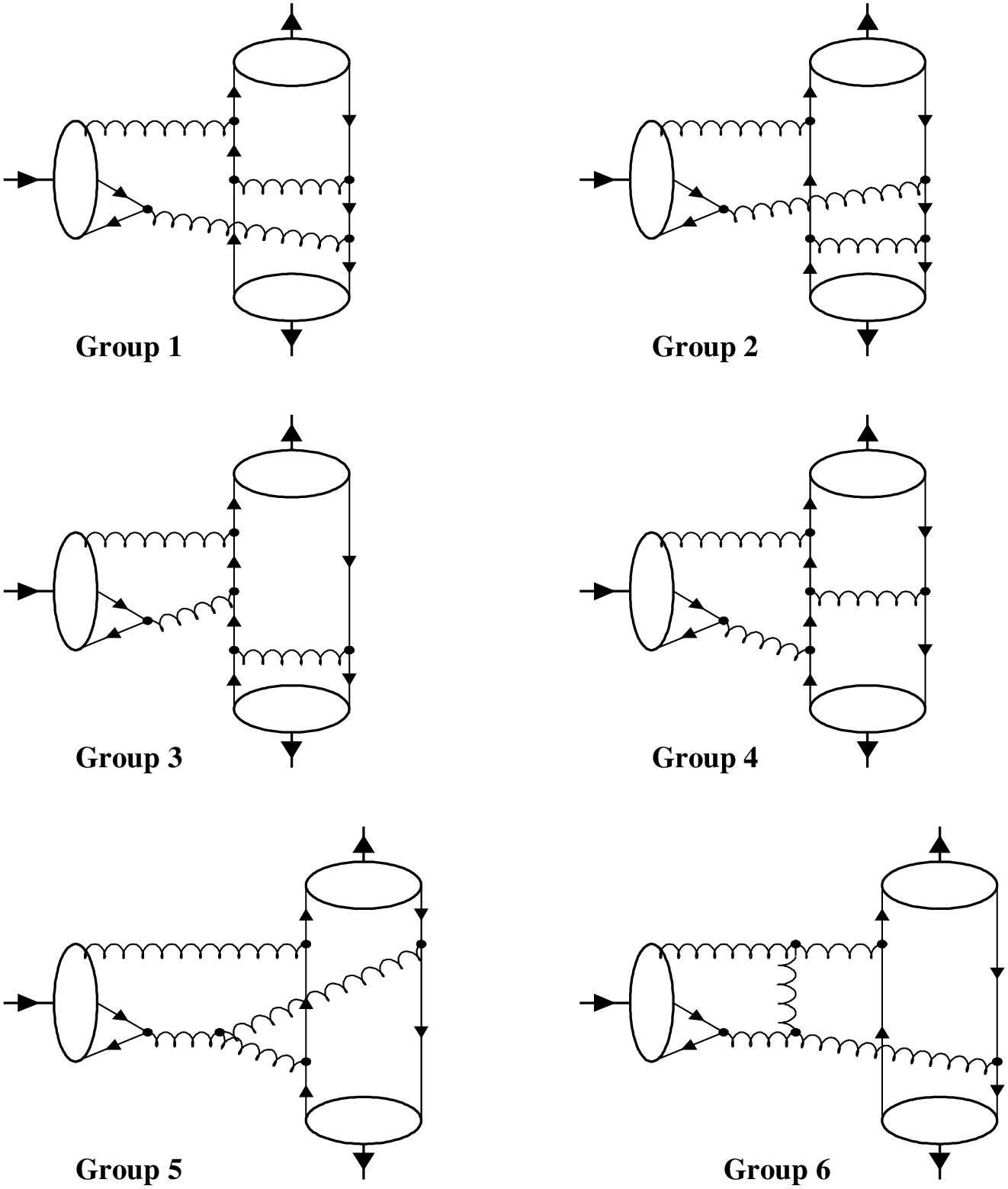,%
       bbllx=10pt,bblly=335pt,bburx=580pt,bbury=810pt,%
       width=7.5cm,clip=} }
   \put(85,30){\psfig{figure=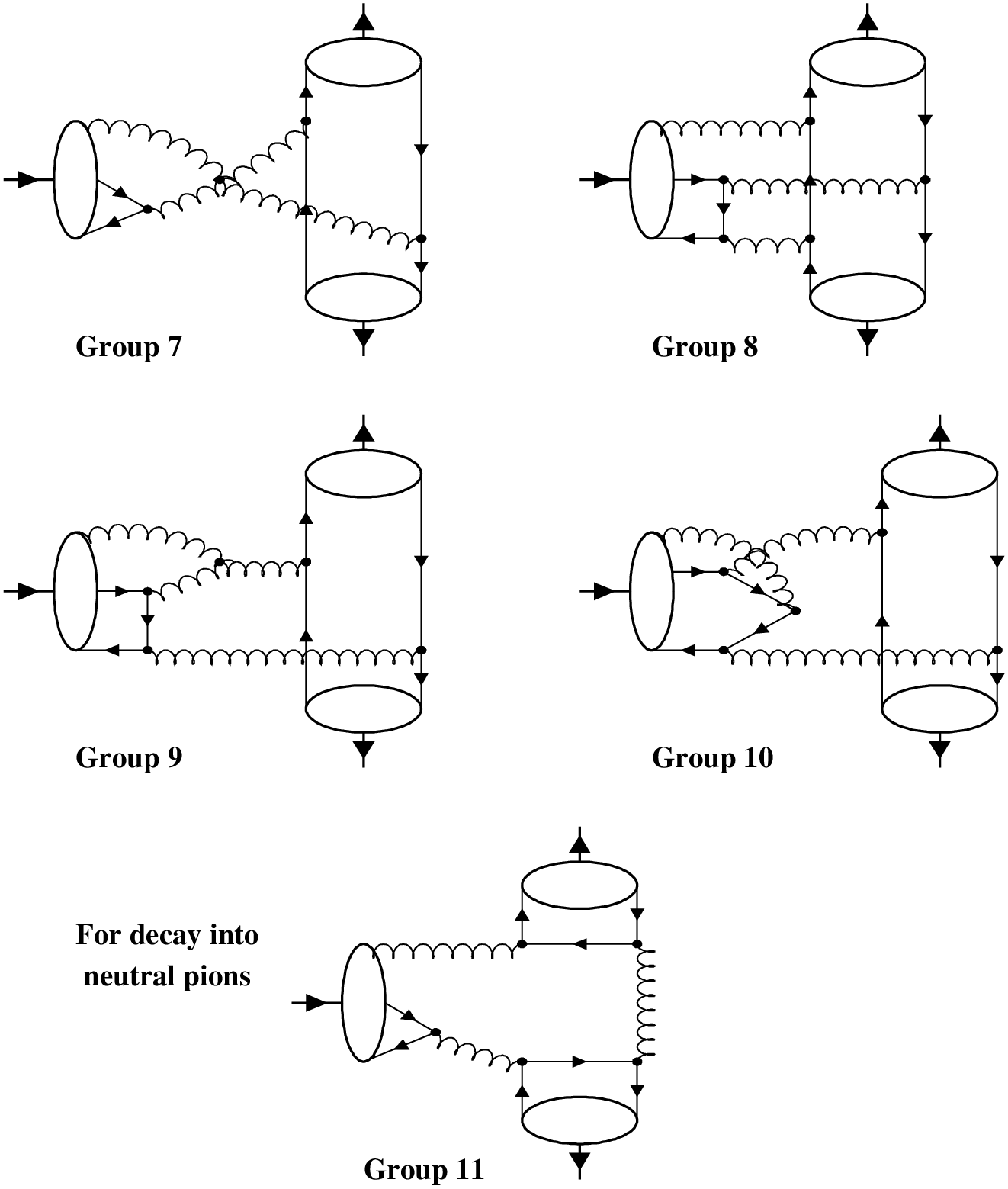,%
       bbllx=10pt,bblly=335pt,bburx=580pt,bbury=810pt,%
       width=7.5cm,clip=} }
 \end{picture}
 \caption{Representatives of the various groups of colour-octet decay
  graphs.
  \label{fig:col8graphs} }
\end{figure}

Our solution to colour conservation is the following:
We attach the gluon of the $|\c\cbar_8\g\rangle$ state 
(in all possible ways) to the hard process leading to the 
Feynman diagrams shown in Fig.~\ref{fig:col8graphs}. 
%%%%%%%%%%%%%%%%%%%%%%%%%%%%%%%%%%%%%%%%%%%%%%%%%%%%%%%%%%%%%%%%%%%%%%%
We do, however, realize that the virtualities
of the quark propagators to which the soft gluon couples are typically
(much) smaller than the virtualities associated with the hard gluons. 
Therefore we treat the coupling of the soft gluon as a parameter 
separate from $\alpha_s$ and denote it by $\alpha_s^{\rm soft}$. 
It will turn out that the final result depends on 
just a single non-perturbative parameter, in which $\alpha_s^{\rm soft}$
appears as a factor.

It is important to realize that in the $|\c\cbar_8\g\rangle$
Fock state not only the $\c\cbar$ pair is in a colour-octet state, but 
also the three particles, $\c$, $\cbar$ and $\g$, are in an $S$-state. The 
covariant spin wave functions we are using for the $\chi_{\c J}$ read
\begin{equation}
  \label{csw8}
  S_{0\nu}^{(8)}\,=\, \frac{1}{\sqrt{6}}\,( p\sla\,+\, M_0)\, 
               ( p_{\nu}/M_0 -\gamma_{\nu} ), \quad
  S_{2\nu}^{(8)}\,=\, \frac{1}{\sqrt{2}}\,( p\sla\,+\,
  M_2)\,\varepsilon_{\mu\nu} \gamma^{\mu}. 
\end{equation}
The colour-octet component of the $\chi_{\c J}$ is given by 
\begin{equation}
  \label{coc}
  |\chi_{\c J}^{(8)} \rangle\, = \, \frac{t^a_{\c\cbar}}{2} \,
                      f_J^{(8)}\,\int \d z_1\d z_2
                      \Phi_J^{(8)}(z_1,z_2,z_3) \, S_{J\nu}^{(8)},
\end{equation}
where $t=\lambda/2$ is the Gell-Mann colour matrix and $a$ the colour
of the gluon. Since orbital angular momenta are not
involved, the transverse degrees of freedom are already integrated
over. Therefore, we only have to operate with a \da\ that is, as usual, 
subject to the condition $\int \d z_1\d z_2 \Phi_J^{(8)}=1$, and 
an octet decay constant $f_J^{(8)}$. In the following, we will make 
the plausible assumption that the colour-octet $\chi_{\c J}$ states 
differ only by their spin wave functions, i.e.\ the \da s as well as 
the decay constants are assumed to be the same for the two $\chi_{\c J}$
states. 

Usually higher Fock state contributions to
exclusive reactions are suppressed by powers of $1/Q^2$
\cite{Brodsky80} where $Q=\mc$ in our case. However, for the decays of
$P$-wave charmonia this suppression does not appear as a simple dimensional
argument reveals: the colour-singlet and octet contributions to the decay
amplitude defined in (\ref{width}) behave as 
\begin{equation}
  \label{powers}
  M^{(c)}_J \sim \fpi^2 f^{(c)}_J \mc^{-n_{c}}.
\end{equation}
The singlet decay constant, $f^{(1)}_J$, represents the derivative of a
two-particle coordinate space wave function at the origin. Hence it is
of dimension GeV$^2$. The octet decay constant, $f^{(8)}_J$, as
a three-particle coordinate space wave function at the origin, is also
of dimension GeV$^2$. Since, according to (\ref{width}), $M^{(c)}_J$
is of dimension GeV, $n_{c}=3$ in both cases.
We note that the $\chi_{\c J}$ decay constants may also 
depend\footnote{In the $\chi_{c0}$ case for instance the spin wave 
for the $\c\cbar$ Fock state reads $ S_0^{(1)}\, =\, \frac{1}{\sqrt{2}} 
\left [p\sla\,+\, M_0\,+\,2K\sla \right ]\, K\sla$. 
The decay constant is then defined as 
$f_0^{(1)} = \frac{4\pi}{3} \int \frac {\d k k^4}{16\pi^3 M_0}
\Psi_0^{(1)}(k) = -\imath |R'_P(0)|/\sqrt{16\pi \mc}$.} on $\mc$.

We estimate the colour-octet contributions to the $\chi_{\c J}$ decays by
calculating the set of Feynman graphs of which representatives are
shown in Fig.~\ref{fig:col8graphs}. The graphs of group 11 only 
contribute to decays into $\pi^0\pi^0$. In order to simplify matters 
we carry through the calculation in the collinear approximation
and assume a $\delta$-function-like $\c\cbar\g$ distribution amplitude
($z_1=z_2=(1-z)/2$; $z_3=z \simeq (1-2\mc/\Mc) \simeq 0.15$, where
$\Mc$ is the average charmonium mass and $z_3$ is the momentum fraction 
carried by the constituent gluon). Most of the details of the calculation, 
as well as the application of the modified HSA, are left to a forthcoming
paper \cite{Bol96a}. We note that in some of the graphs shown in 
Fig.~\ref{fig:col8graphs} the virtuality of the quark propagator adjacent 
to the $\chi_{\c J}$ constituent gluon is small. 
It is these diagrams that constitute, to leading order in $\alpha_s$
and $z$, 
the higher Fock state $|\q\qbar\g\rangle$ of the pion. 
However, for arbitrary $z$, these diagrams alone do not lead 
to a gauge-invariant amplitude.    

In the calculation of the colour-octet contribution, a number of 
singular integrals appear as a consequence of the collinear
approximation. If the transverse momenta are kept, all integrals are
well defined. The difficulty arises from the propagator 
$D_{ca}=[(z-x)(z-y) +\imath\epsilon]^{-1}$ of the gluon exchanged between 
the light quarks for which the usual
$\imath\epsilon$ prescription fails. In order to regularize these
integrals we replace $D_{ca}$ by 
\begin{equation}
  \label{prop}
  D = [(z-x)(z-y) + \varrho^2 +\imath\epsilon]^{-1}.
\end{equation}
The integrals can then be worked out straightforwardly \cite{Far89}. The
regulator $\varrho$ represents a mean transverse momentum of the quarks inside
the pions. Indeed,
\begin{equation}
  \label{regu}
  \varrho^2 = \langle k^2_{\perp}\rangle / (4\mc^2) \ .
\end{equation}
We checked that for appropriate values of $\varrho$ this regularisation
recipe provides, to a reasonable approximation, similar 
numerical results as one
obtains when all transverse quark momenta are kept and, weighted by
Gaussian $k_{\perp}$-dependences, integrated over. 

The final results for the colour-octet contribution to the decay amplitudes
can again be written in the form (\ref{decayone}) with 
\begin{equation}
  \label{octet}
             a_J^{(8)}=\kappa \tilde{a}_J(z,\varrho)\,,\quad
             b_J^{(8)}=\kappa \tilde{b}_J(z,\varrho)\,,\quad
             c_J^{(8)}=\kappa \tilde{c}_J(z,\varrho)
\ ,
\end{equation}
where $\tilde{a}_J$, $\tilde{b}_J$ and $\tilde{c}_J$ are only weakly 
dependent on
the exact values of the parameters $z$ and $\varrho$. Thus, it appears
reasonable to take $\kappa$, defined by
\begin{equation}
  \label{kappa}
  \kappa = \sqrt{\alphas^{\rm soft}} f^{(8)}_0/\varrho^2
\end{equation}
as the only fit parameter. Evaluating the colour
octet contribution from the asymptotic $\pi$ wave function ($B_2=0$) 
with $z=0.15$, $\varrho$ about $0.1$, which corresponds to some typical
transverse momentum of $300\,$MeV, and, guided by our
results for the singlet contribution obtained within the modified HSA,
$\alphas(\mu_R^{excl}) =0.45$, we find 
\begin{equation}
        \tilde{a}_0  =  (151.1 +\imath 25.9)\,{\rm GeV}^{-2}
   \quad , \quad  
        \tilde{a}_2  =  (156.1 +\imath  6.6)\,{\rm GeV}^{-2} 
\ .
%\end{tabular}
   \label{coloctAS1}
\end{equation}
The quantities $\tilde{b}_J$ and $\tilde{c}_J$, necessary for 
$B_2 \neq 0$ case, will be given in \cite{Bol96a}.

The $\chi_{\c J}$ decay widths into pions are now obtained by
adding coherently the respective colour-singlet and colour-octet 
contributions. Since both contributions have the same scaling with
$1/c$
they have to be evaluated in the same scheme, i.e.\ the colour-singlet 
result in the collinear approximation has to be used.
The four widths, $\chi_{\c J} \rightarrow \pi^+\, \pi^-$, 
$\pi^0\,\pi^0$ for $J=0$, $2$ are given in terms of the single 
non-perturbative factor $\kappa$. A fit yields 
$\kappa = 0.161\,$GeV$^2$ and the individual widths are
shown in tables~\ref{tab:results} and~\ref{tab:pionzero}.
%%%%%%%%%%%%%%%%%%%%%%%%%%%%%%%%%%%%%%%%%%%%%%%%%%%%%%%%%%%%%%%%%%%%%%%%%
\begin{table}
\begin{center}
\begin{tabular}{|c|r|r|r|r|} \hline
  & \multicolumn{2}{|c|}{$\Gamma(\chi_{\c J} \to \pi^0\pi^0)\,$[keV]  }
  & \multicolumn{2}{|c|}{$\BR(\chi_{\c J} \to \pi^0\pi^0)\,$[\%]}
\\ \hline
  $B_2 = 0$ & $25.7$ & $1.81$ & $0.18$ & $0.091$
\\ \hline
 Exp.\ PDG \cite{pdg94} & $42 \pm 18$ & $2.2 \pm 0.6$ & 
  $0.31 \pm 0.06$ & $0.110 \pm 0.028$
\\ \hline
\end{tabular}
\caption[dummy]{Decay widths and branching ratios of $\chi_{\c J}
\rightarrow \pi^0 \pi^0$ (colour-octet contributions included; 
$\mc=1.5\,$GeV, $\LQCD=0.2\,$GeV).}
\label{tab:pionzero}
\end{center}
\end{table}
We present also predictions for the respective branching ratios using the
experimentally measured total widths.
As shown in tables~\ref{tab:results} and~\ref{tab:pionzero} the 
inclusion of the colour-octet mechanism brings predictions and
data in generally good agreement.

Finally we address the question whether the value found for the
parameter $\kappa$ is sensible. For this purpose we estimate the 
probability for the $\chi_{\c 0}$ to be in the colour-octet Fock state from
a plausible wave function:
\begin{eqnarray}
  \label{bsw}
  \Psi_{0}^{(8)}(z_i,{\bf k}_{\perp i}) = N z_1 z_2 z_3^2 &&\,
      \exp\left\{-2a_{\chi}^{2}\mc^2\,
                \left[ (z_3-z)^2 + (z_1-z_2)^2 \right]\right\}\nonumber\\
                &&\times \exp\left\{-a_{\chi} \sum k^2_{\perp i}\right\}
\ .
\end{eqnarray}
This ansatz combines the known asymptotic   
behaviour of a \da\ for a $\q\qbar\g$ Fock state \cite{CZ84r}
with mass-dependent exponentials and a Gaussian $k_{\perp}$-dependence
in analogy to the Bauer-Stech-Wirbel parametrization of 
charmed-meson wave functions \cite{Bsw85}. The mass-dependent exponentials 
guarantee a pronounced peak of the \da\ at $z_1 \simeq z_2 \simeq \mc/\Mc$. 
The $\delta$-function-like \da\ used in the estimate of the 
colour-octet contribution appears as the peaking approximation to 
this function. Since the $\c\cbar\g$ Fock state is an $S$-wave state we 
assume its radius to be equal to that of the $S$-state charmonia, namely 
$0.42\;{\rm fm}$ \cite{BuT81}. In this case the
oscillator parameter $a_{\chi}$ takes the value $1.23$ GeV$^{-1}$. 
The probability of the colour-octet Fock state is 
\begin{equation}
  \label{ocprob}
  P_{\c\cbar\g} = \left ( f_0^{(8)}/2.1\times 10^{-3} {\rm GeV}^{2}
                                                               \right )^2
\ .
\end{equation}
This relation is obtained with $z = 0.15$; it is however practically
independent of the exact value of $z$. 
Taking $\alpha_s^{\rm soft} \simeq \pi$ we obtain 
from (\ref{kappa}) $f_0^{(8)}\simeq 0.9 \times 10^{-3}\,{\rm GeV}^2$
and $P_{\c\cbar \g} \simeq 19\%$. Hence, the probability for the
$\chi_{\c J}$ to be in the $\c\cbar\g$ state comes out much more
reasonable than one might have hoped for.

In summary, we have shown that the constraints on the pion-\da\ 
from $F_{\pi\gamma}(Q^2)$ data lead to predictions for
the exclusive decay $\chi_{\c J} \rightarrow \pi\pi$ that fall
well (a factor of ten) below the data when restricting to the dominant, 
colour-singlet $|\c\cbar({}^3P_J)\rangle$ Fock state. The discrepancies 
persist after the inclusion of transverse degrees of freedom and Sudakov 
suppressions in the modified HSA approach. The deviations are, in fact, 
so large that they cannot sensibly be explained by
higher-order perturbative QCD ($\alpha_s$) or genuine higher-twist 
($1/\mc$) corrections. 

We have shown that contributions to the exclusive decay from the 
$O(v)$-suppressed Fock component $|\c\cbar\g\rangle$ are actually of 
the same power in $1/\mc$ as the conventional colour-singlet contribution. 
Predictions including this colour-octet mechanism basically require only  
a single additional non-perturbative parameter. With quite a reasonable 
value for it, we obtain decay widths and branching ratios for the
exclusive decays $\chi_{\c J}\rightarrow \pi^+\pi^-$ and $\pi^0\pi^0$
which are in good agreement with data.  

We do not expect a significant colour-octet contribution 
to $\JP$ (and $\Upsilon$) decay into pions, since contributions 
from the higher-Fock state $|\c\cbar({}^3P_J)\g\rangle$ are truly
suppressed by $1/\mc^2$. The solution to a similar discrepancy between
data and theory for $\JP \rightarrow \pi^+\pi^-$ is therefore still
open and needs further study. 
Our findings for exclusive $\chi_{\c J}$ apply to 
$P$-wave bottomonium decays $\chi_{\b J} \rightarrow \pi\pi$ as well:
a large fraction of the width will originate from the $\b\bbar\g$ state.
One may speculate that the colour-octet mechanism is relevant 
to other exclusive decays as well, e.g.\ $B\rightarrow \chi_{\c J} K$.
Work along these lines is in progress.\hfill\\[1ex]

\noindent
{\it Acknowledgements:} One of us (G.A.S) wants to thank the Theory
Group of SLAC for hospitality where part of the work was completed
and S.\ Brodsky for enlightening discussions.
%%%%%%%%%%%%%%%%%%%%%%%%%%%%%%%%%%%%%%%%%%%%%%%%%%%%%%%%%%%%%%%%%%%%

\end{document}